\documentstyle[prl,aps,twocolumn]{revtex}

\begin{document}
\input epsf
\draft
\tighten

\def\bk {{ \mathbf{k} }}
\def\bkappa   { \mbox{\boldmath ${\kappa}$}   }
\hyphenation{cu-mu-lant}
\hyphenation{cu-mu-lants}
\hyphenation{azi-mu-thal}


\preprint{mpi-pks/9712007, STPHY 27/97, HEPHY-PUB 676/97}

\title{Analytic multivariate generating function for random
multiplicative cascade processes\cite{PCnote}}

\author{Martin Greiner$^{1}$, Hans C.\ Eggers$^{2}$ and Peter Lipa$^3$}
\address{$^1$Max-Planck-Institut f\"ur Physik komplexer Systeme, 
         N\"othnitzer Str.\ 38, D--01187 Dresden, Germany}
\address{$^2$Department of Physics, University of Stellenbosch,
         7600 Stellenbosch, South Africa}
\address{$^3$Institut f\"ur Hochenergiephysik,
         \"Osterreichische Akademie der Wissenschaften, 
         Nikolsdorfergasse 18, A--1050 Vienna, Austria}

\date{12 December 1997 {\bf  chao-dyn/9804024} }

\maketitle

\begin{abstract}
We have found an analytic expression for the multivariate
generating function governing all $n$-point statistics of
random multiplicative cascade processes. The variable appropriate
for this generating function is the logarithm of the energy density,
$\ln\epsilon$, rather than $\epsilon$ itself. All cumulant
statistics become sums over derivatives of
``branching generating functions'' which are Laplace transforms
of the splitting functions and completely determine
the cascade process. We show that the branching generating function
is a generalization of the multifractal mass exponents.
Two simple models from fully developed turbulence illustrate 
the new formalism.
\end{abstract}


\narrowtext


Multifractals have become a popular tool especially in fully developed 
turbulence to analyze the intermittent fluctuations
occurring in the energy dissipation field \cite{MEN91,FRI95}. 
They have also been applied to characterize the phase space
structure of strange attractors in chaotic dynamical systems 
\cite{HAL86}, diffusion limited aggregation \cite{ARG90}, 
high-energetic multiparticle dynamics \cite{WOL96}, 
geophysics and self-organized criticality \cite{GAR96}, 
to name just a few.

Among the simplest
examples of multifractal processes are random multiplicative
cascade processes in general and the multiplicative binomial
process \cite{FED88,MEN87} in particular. 
Both the multifractal approach and large deviation theory
analyse these processes in terms of one-point statistics.
However, because different cascade processes like, for example, the 
$p$-model \cite{MEN87} and the $\alpha$-model \cite{SCH85}
are indistinguishable in this framework,
such one-point statistics appear to be too restrictive.
In order to discriminate among these and other suggested models, 
a generalisation of the multifractal approach 
to $n$-point statistics is thus called for.
The analytic multivariate generating function presented in this letter 
is, we believe, the appropriate generalisation.

Two-point statistics of random multiplicative cascade
processes were previously discussed in Refs.\ \cite{CAT87,MEN90}.
An approach to calculate the $n$-point statistics, or in 
other words the spatial correlations to arbitrary order,
within these models has been presented in 
Refs.\ \cite{GRE95,GRE96}: there, the multivariate generating function 
of the spatial correlations was constructed iteratively
from a backward evolution equation, leading to a recursive
derivation of spatial correlations. While in the latter approach, 
models like the $p$- and $\alpha$-model become clearly distinguishable,
the recursive structure of the multivariate generating function 
appears from a mathematical perspective to be less than elegant:
an analytic solution is much more desireable. The realization
of this goal is the subject of this letter.

In order to derive the analytic expression for the multivariate
generating function of random multiplicative binary cascade
processes, we first need to find the appropriate variables
and to introduce a convenient labelling. The cascade prescription
is as follows: 
A given initial interval of length 1 and unit energy density 
successively splits into 2, 4, 8, \ldots subintervals (``bins''). 
For the latter, we employ a binary labeling: after $j$ 
cascade steps (interval splittings), a specific subinterval
is characterized by the sequence $k_1 k_2 \cdots k_j$ with
each $k_i$ being either 0 or 1. The bin with the
label $00\cdots 0$ is the leftmost one, whereas the label
$11\cdots 1$ belongs to the bin on the far right.
The energy density $\epsilon_{k_1 \cdots k_j}^{(j)}$ 
belonging to a $j$-th generation bin is redistributed
nonuniformly onto the two intervals of the next generation:
$\epsilon_{k_1 \cdots k_j0}^{(j+1)} = q_{k_1 \cdots k_j 0}^{(j+1)}
\epsilon_{k_1 \cdots k_j}^{(j)}$, 
and 
$\epsilon_{k_1 \cdots k_j1}^{(j+1)} = q_{k_1 \cdots k_j 1}^{(j+1)}
\epsilon_{k_1 \cdots k_j}^{(j)}$.
The two multipliers $q_{k_1 \cdots k_j 0}$ and $q_{k_1 \cdots k_j 1}$
are drawn from a probability density
$p(q_{k_1 \cdots k_j 0}, q_{k_1 \cdots k_j 1})$
which we call the splitting function.

After $J$ cascade steps, the energy density belonging to
bin $(k_1 k_2 \cdots k_J)$ is given by the product of 
multipliers over all previous generations
\begin{equation}
\label{eins}
\epsilon_{k_1 \cdots k_J}^{(J)} = 
q_{k_1}^{(1)} 
q_{k_1 k_2}^{(2)} \cdots
q_{k_1 \cdots k_J}^{(J)}.
\end{equation}
For the logarithm of the energy density this relation becomes
of course additive; this observation is crucial for the subsequent
derivation.

The structure of a specific random multiplicative cascade 
model is completely determined by the splitting function
$p(q_0, q_1)$ which determines the distribution of energy
during each split. 
Hence, $p(q_0,q_1)$ also completely determines the multivariate
statistics of a given cascade. The joint probability density 
$\tilde{p}(\ln\epsilon_{0\cdots 00}^{(J)}, \ldots, 
           \ln\epsilon_{1\cdots 11}^{(J)})$
to find at the same time the logarithm of 
$\epsilon_{0\cdots 00}^{(J)}$ in the bin with label $(0\cdots 00)$, 
the value $\ln\epsilon_{0\cdots 01}^{(J)}$ in bin
$(0\cdots 01)$, \ldots, and
$\ln\epsilon_{1\cdots 11}^{(J)}$ in bin $(1\cdots 11)$
can therefore be expressed fully in terms of the splitting
functions at each branching: 
\FL
\begin{eqnarray}
\label{drei}
&&
\tilde{p}(\ln\epsilon_{0\cdots 0}^{(J)}, \ldots, 
          \ln\epsilon_{1\cdots 1}^{(J)})
=
\nonumber \\
&& 
\int \Bigl[ \prod_{j=1}^J 
     \prod_{  {k_1, \ldots, \atop k_{j-1}=0 }  }^1
            dq_{k_1 \cdots k_{j-1} 0}^{(j)}
            dq_{k_1 \cdots k_{j-1} 1}^{(j)}
            p ( q_{k_1 \cdots k_{j-1} 0}^{(j)},
                q_{k_1 \cdots k_{j-1} 1}^{(j)}  )
     \Bigr] \,
\nonumber \\
&& \qquad \times
\Biggl[
            \prod_{  {k_1, \ldots, \atop k_{J}=0 }  }^1
    \delta \Bigl( \ln\epsilon_{k_1 \cdots k_J}^{(J)}
                -  \sum_{j=1}^J \ln q_{k_1 \cdots k_j}^{(j)}
           \Bigr)
\Biggr].
\end{eqnarray}
This probability density can be converted into a 
multivariate generating function for cumulants in
$\ln\epsilon$. With the definition
\FL
\begin{eqnarray}
\label{vier}
&&
K [ \lambda_{0\cdots 0}^{(J)} , \ldots, \lambda_{1\cdots 1}^{(J)} ]
= 
\ln \left\langle
\exp \Biggl( 
    \sum_{k_1,\ldots, \atop k_J = 0}^1 
      \lambda_{k_1\cdots k_J}^{(J)} 
      \ln \epsilon_{k_1\cdots k_J}^{(J)} 
     \Biggr)
\right\rangle
\nonumber \\
&& =
\ln\Biggl[ \int \Bigl(
            \prod_{  {k_1, \ldots, \atop k_{J}=0 }  }^1
            d(\ln \epsilon_{k_1\cdots k_J}^{(J)} )
                \Bigr) \;
p(\ln\epsilon_{0\cdots 0}^{(J)}, \ldots, \ln\epsilon_{1\cdots 1}^{(J)})
\Biggr.
\nonumber \\
&&
\Biggl.\qquad
\times\exp \Biggl( 
    \sum_{k_1,\ldots, \atop k_J = 0}^1 
      \lambda_{k_1\cdots k_J}^{(J)} 
      \ln \epsilon_{k_1\cdots k_J}^{(J)} 
     \Biggr)
\Biggr]
\end{eqnarray}
we find, after defining
\begin{equation}
\label{sechs}
\lambda_{k_1\cdots k_j}^{(j)} 
= \sum_{k_{j+1}, \ldots, k_J = 0}^1 
\lambda_{k_1\cdots k_j k_{j+1} \cdots k_J}^{(J)} \,,
\end{equation}
rearranging the terms in the exponent of (\ref{vier}),
namely
\FL
\begin{eqnarray}
\label{hceb}
&&
\sum_{k_1, \ldots, k_J} \lambda_{k_1 \cdots k_J}^{(J)}
                    \ln\epsilon_{k_1 \cdots k_J}^{(J)}
= 
\sum_{j=1}^J 
\sum_{k_1, \ldots, k_j}
          \lambda_{k_1 \cdots k_j}^{(j)}
            \ln q_{k_1 \cdots k_j}^{(j)}
\\
&=&
\sum_{j=1}^J \sum_{{k_1,\ldots, \atop k_{j-1}}}
\left(
  \lambda_{k_1 \cdots k_{j-1}0}^{(j)} \ln q_{k_1 \cdots k_{j-1}0}^{(j)}
+ \lambda_{k_1 \cdots k_{j-1}1}^{(j)} \ln q_{k_1 \cdots k_{j-1}1}^{(j)}
\right)
\,,
\nonumber
\end{eqnarray}
and inserting (\ref{drei}) into (\ref{vier}),
that the $2^J$-fold integral factorizes, so that
\begin{eqnarray}
\label{vieraa}
K [ \lambda_{0\cdots 0}^{(J)} , \ldots, \lambda_{1\cdots 1}^{(J)} ]
&& =
\sum_{j=1}^J \sum_{k_1,\ldots, \atop k_{j-1} = 0}^1
Q[ \lambda_{k_1\cdots k_{j-1} 0}^{(j)},
   \lambda_{k_1\cdots k_{j-1} 1}^{(j)}],
\end{eqnarray}
where
\begin{equation}
\label{fuenf}
Q[\lambda_0, \lambda_1] 
= \ln
\Bigl[
   \int dq_0\, dq_1 \,  p(q_0, q_1) 
   e^{\lambda_0 \ln q_0 + \lambda_1 \ln q_1}
\Bigr] 
\end{equation}
is the ``branching generating function'' (b.g.f.)
for cumulants, 
which governs the behavior of the entire cascade.
Eqs.\ (\ref{vieraa})--(\ref{fuenf}) 
represent the long-sought analytic expression
for multiplicative cascades: because
the multivariate cumulant generating function $K$
is the sum of all branching
generating functions $Q$, one for every branching, each
b.g.f.\ can be solved separately and analytically.

Multivariate cumulants likewise become sums
over cumulants of the individual branching points.
Using the notation $\bkappa \equiv (k_1 \cdots k_J)$ in the
subscripts, the multivariate cumulant of order $n$ is
found through
\begin{eqnarray}
\label{cumdev}
C_{\bkappa_1, \ldots, \bkappa_n} 
&=&
\Bigr.
{\partial^n K[\lambda^{(J)}] \over
\partial\lambda_{\bkappa_1} \cdots
\partial\lambda_{\bkappa_n}
} 
\Bigl|_{\lambda^{(J)} = 0}.
\end{eqnarray}
The first three multivariate cumulants are
\begin{eqnarray}
\label{cumls}
C_{\bkappa_1} 
&=&
    \langle \ln\epsilon_{\bkappa_1} \rangle_c
=   \langle \ln\epsilon_{\bkappa_1} \rangle ,
\\
C_{\bkappa_1, \bkappa_2}
&=& \langle \ln\epsilon_{\bkappa_1} 
            \ln\epsilon_{\bkappa_2}
    \rangle_c
\nonumber \\
&=& \langle \ln\epsilon_{\bkappa_1} 
            \ln\epsilon_{\bkappa_2}
    \rangle
-   \langle \ln\epsilon_{\bkappa_1} \rangle
    \langle \ln\epsilon_{\bkappa_2} \rangle,
\\
C_{\bkappa_1, \bkappa_2, \bkappa_3}
&=& \langle \ln\epsilon_{\bkappa_1} 
            \ln\epsilon_{\bkappa_2}
            \ln\epsilon_{\bkappa_3}
    \rangle_c
\nonumber \\
&=& \langle \ln\epsilon_{\bkappa_1} 
            \ln\epsilon_{\bkappa_2}
            \ln\epsilon_{\bkappa_3}
    \rangle
\nonumber \\
&& - \sum_{\rm perm}
    \langle \ln\epsilon_{\bkappa_1} 
            \ln\epsilon_{\bkappa_2}
    \rangle
    \langle \ln\epsilon_{\bkappa_3} \rangle
\nonumber \\
&& + 2
    \langle \ln\epsilon_{\bkappa_1} \rangle
    \langle \ln\epsilon_{\bkappa_2} \rangle
    \langle \ln\epsilon_{\bkappa_3} \rangle.
\end{eqnarray}
These and higher-order cumulants can be found through
appropriate derivatives of the b.g.f.\ using (\ref{vieraa}). 
If, for example,
the same b.g.f.\ is used for all branchings, and the
splitting function is symmetric, $p(q_0, q_1) = p(q_1, q_0)$,
the two-point cumulant is
\FL
\begin{eqnarray}
\label{sieben}
C_{\bkappa,\bkappa^\prime}
&=& 
\Biggl(
 \sum_{j=1}^J \delta_{k_1 k_1^\prime} \cdots \delta_{k_j k_j^\prime}
\Biggr)
\left.
{
\partial^2 Q[\lambda_0, \lambda_1] 
\over
\partial \lambda_0^2 
}
\right|_{\lambda = 0}
\\
&&
+ \Bigl( 1 -  \delta_{k_1 k_1^\prime} \cdots \delta_{k_J k_J^\prime}
  \Bigr)
\left.
{
\partial^2 Q[\lambda_0, \lambda_1] 
\over
\partial \lambda_0 \partial \lambda_1
}
\right|_{\lambda = 0}.
\nonumber 
\end{eqnarray}
For the $p$-model with a splitting function 
$p(q_0, q_1) = \textstyle{1\over 2} 
[ \delta(q_0 - (1+\alpha)) + \delta(q_0 - (1-\alpha))]
  \delta(q_0 + q_1 - 2)$,
the second-order cumulant correlation density is shown in Fig.\ 1.

We should point out the scope and limitations of the solution
(\ref{vieraa}). Clearly, the generating function in terms of
$\ln\epsilon$ is applicable to any functional form of the
splitting function or b.g.f. It does not depend on the number of
branches either: trivariate or even higher-variate splitting
functions can be implemented. Due to the additive nature of
the b.g.f.'s, the splitting functions can differ from
generation to generation and even from branch to branch.
The only (and important) precondition for the applicability
of Eq.\ (\ref{vieraa}) is that the splitting variables $q$
of every branching must be independent of all others\footnote{
Nelkin and Stolovitzky \cite{NEL96} argue that the dependence of
experimental distributions of splitting variables
(multiplier distributions) on the position of the subinterval
\cite{PED96} suggests that the splitting variables are in fact
not statistically independent; see also Ref.\ \cite{SRE95}.
The effects of this deviation from statistical independence 
on the present formalism, and to what extent it matters,
remain to be investigated.
}.

\mbox{ }\par\vspace*{-30mm}\par
\begin{center}
\epsfysize=127mm
\epsfbox{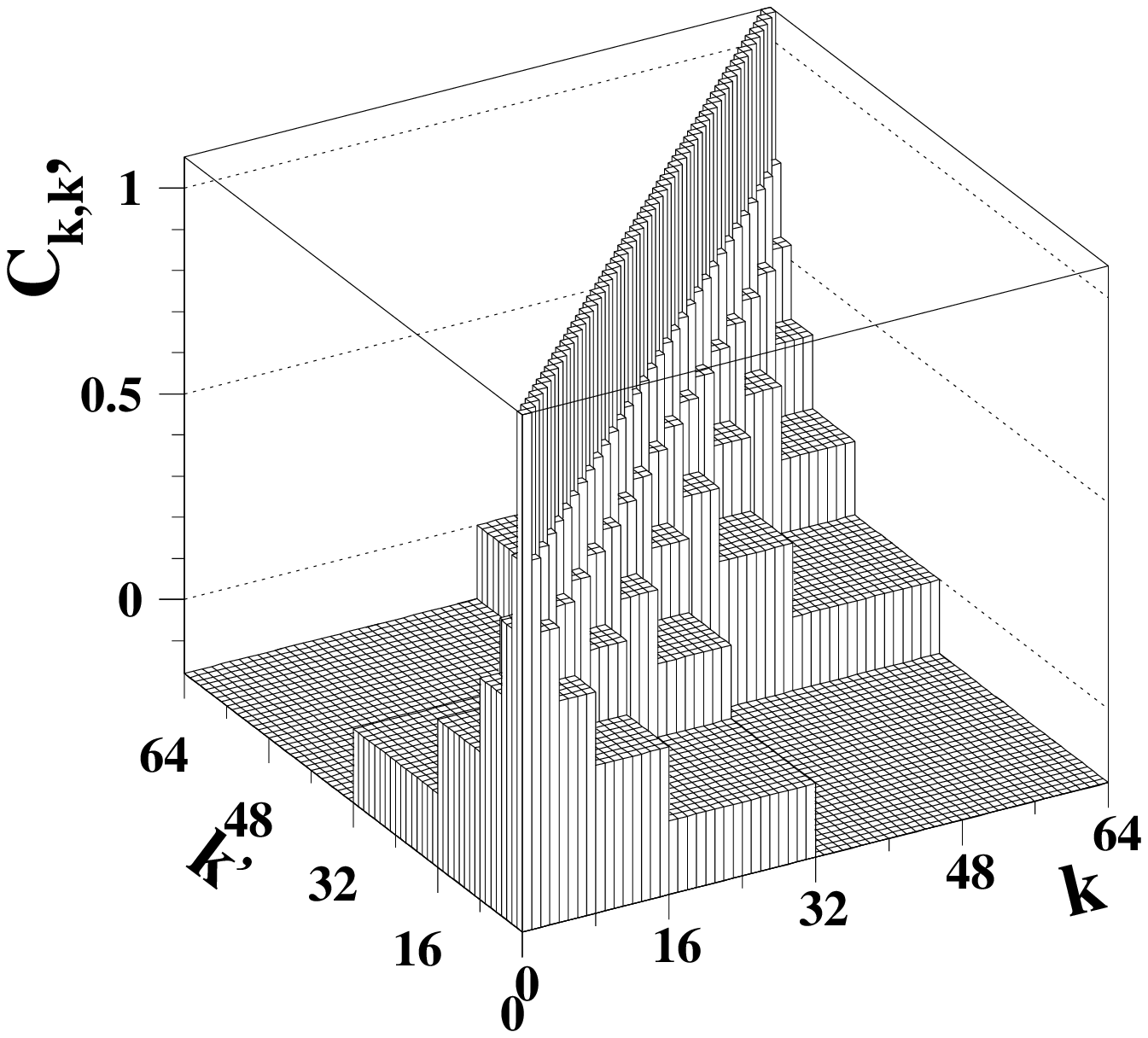}
\end{center}
\vspace*{-39mm}\par
{\noindent\bf Figure 1:}\\
Second-order cumulant
$C_{k,k^\prime} \equiv 
C_{(k_1 \cdots k_J),(k_1^\prime \cdots k_J^\prime)}$
for the $p$-model, where 
$k = 1 + \sum_{j=1}^J k_j 2^{J-j}$ and \\
$k^\prime = 1 + \sum_{j=1}^J k_j^\prime 2^{J-j}$.
Parameters have been set to $J=6$ and $\alpha = 0.4$.
\\

Moments of the $J$-generation cascade are found directly from
the moment generating function $Z[\lambda^{(J)}] 
= \exp(K[\lambda^{(J)}])$.
{}From Eq.\ (\ref{vieraa}), it is obvious that $Z$ factorizes
into moment generating functions of the individual branchings.

The key input into the analytical expression (\ref{vieraa})
is the branching generating function $Q[\lambda_0, \lambda_1]$
of Eq.\ (\ref{fuenf}). Its properties uniquely fix the spatial 
correlations of the binary random multiplicative cascade model.
For the $p$-model, we get
\begin{eqnarray}
\label{acht}
Q[\lambda_0, \lambda_1]_{p{\rm -model}}
&=&
\textstyle{1\over 2}  (\lambda_0 + \lambda_1) \ln(1 - \alpha^2)
 \\
&+& 
\ln \left\{
  \cosh\left[ {1\over 2}\, (\lambda_0 - \lambda_1)
        \ln\left( 1 + \alpha \over 1 - \alpha \right)
       \right]
     \right\}.
\nonumber
\end{eqnarray}
The symmetric $\alpha$-model is similar to the $p$-model
except that it does not conserve energy in a cascade
splitting. Given its splitting function,
$p(q_0, q_1) = \textstyle{1\over 4} \prod_{k=0}^1
\left[ \delta(q_k - (1+\alpha)) + \delta(q_k - (1-\alpha)) \right]$,
its b.g.f.\ then reads
\begin{eqnarray}
\label{neun}
Q[\lambda_0, \lambda_1]_{\alpha{\rm -model}}
&=&
\textstyle{1\over 2}  (\lambda_0 + \lambda_1) \ln(1 - \alpha^2)
 \\
&+& 
\ln \left\{
  \cosh\left[ {1\over 2} \lambda_0
        \ln\left( 1 + \alpha \over 1 - \alpha \right)
       \right]
     \right\}
\nonumber \\
&+& 
\ln \left\{
  \cosh\left[ {1\over 2} \lambda_1
        \ln\left( 1 + \alpha \over 1 - \alpha \right)
       \right]
     \right\}
\nonumber
\end{eqnarray}
which clearly differs from the b.g.f.\  (\ref{acht})
for the $p$-model. Fig.\ 2 compares the two branching
generating functions. Note that for $\lambda_0 = 0$
or $\lambda_1 = 0$ the two expressions (\ref{acht})
and (\ref{neun}) become identical; consequently,
the one-point cumulants
\begin{equation}
\label{zehn}
\left.
{ \partial^n Q[\lambda_0, \lambda_1] \over
\partial \lambda_0^n}
\right|_{\lambda = 0}
=
\left\langle (\ln q_0)^n \right\rangle_c
\end{equation}
of the $p$- and $\alpha$-model are also identical. This is the
reason why, in a multifractal approach,  the two models look the same
asymptotically. To see differences between the two models,
one must  go to the two-point cumulants
\begin{equation}
\label{elf}
\left.
{ \partial^n Q[\lambda_0, \lambda_1] \over
\partial \lambda_0^{n_1} \partial \lambda_1^{n - n_1}}
\right|_{\lambda = 0}
=
\left\langle (\ln q_0)^{n_1} (\ln q_1)^{n - n_1} \right\rangle_c.
\end{equation}
Within the $p$-model, the latter are nonzero for even $n$ and zero for 
odd $n \geq 3$. Within the $\alpha$-model, by contrast, all two-point
cumulants vanish since its splitting function factorizes:
$p(q_0,q_1) = p(q_0) p(q_1)$.
We hence see that the two-point
cumulant moments are sensitive to the violation of energy
conservation in the splitting function.

For a binary random multiplicative cascade process to qualify
as a true multifractal process, the splitting function needs to
conserve energy, i.e.\ $p(q_0, q_1) = p(q_0) \delta(q_0 + q_1 - 2)$;
see Ref.\ \cite{GRE96} for a clarification of this point.
If energy is indeed conserved, it is possible to link 
the multifractal mass exponents $\tau(q)$ to the b.g.f.\
$Q[\lambda_0,\lambda_1]$. Setting $\lambda_1 = 0$ and, for simplicity,
considering the case of a symmetric splitting function
$p(q_0,q_1)=p(q_1,q_0)$, Eq.\ (\ref{fuenf}) 
becomes\footnote{
The univariate version of relation (\ref{zwoelf}) was previously
discussed by Novikov \cite{NOV71} in connection with the statistics
of generalised multipliers, the so-called breakdown coefficients.
}
\begin{eqnarray}
\label{zwoelf}
Q[\lambda_0, \lambda_1 = 0]
&=&
\ln \Bigl[ \int dq_0\, p(q_0) q_0^{\lambda_0} \Bigr]
=
\ln \langle q_0^{\lambda_0}\rangle \nonumber \\
&=& \ln 2
   \Bigl[   (\lambda_0 - 1) + \tau(\lambda_0)   \Bigr];
\end{eqnarray}
note that for clarity we have written
$\tau(\lambda_0)$ instead of the more familiar notation
$\tau(q)$. Hence, for an energy-conserving splitting function,
its multifractal mass exponents follow from
$Q[\lambda_0, \lambda_1]$. The reverse need not be true, though.

Also, for the more general case where energy is not
conserved in the splitting function, the multifractal
mass exponents $\tau(q)$ cannot be deduced
in a clean fashion and contain less information than the b.g.f.
This means that, for binary multiplicative cascades, 
$Q[\lambda_0, \lambda_1]$ 
can be understood as the natural generalization of the multifractal
mass exponents.

For a one-dimensional cut through the three-dimensional energy
dissipation field in fully developed turbulence, it is most
likely that energy is not conserved along the cut. Hence, the
splitting function will not conserve energy and cannot be identified
with the scale-invariant multiplier distributions \cite{SRE95}.
Here it would be interesting to find a procedure to infer
the proper splitting function from data. Given that the experimentally
measurable cumulants in $\ln \epsilon_{k_1 \cdots k_J}^{(J)}$
are $n$-fold derivatives of $Q$, the latter
can in principle be reconstructed from the former.
With the help of Eq.\ (\ref{fuenf}), the b.g.f.\ can then 
be inverted into the splitting function via a two-dimensional 
inverse Laplace transformation,
\FL
\begin{eqnarray}
&&
\int_0^\infty dx \, dy \,
p(2e^{-x}, 2e^{-y}) \,
e^{-(\lambda_0 + 1)x - (\lambda_1 + 1)y }
\nonumber \\
&& \quad 
=
e^{ Q[\lambda_0, \lambda_1] - (\lambda_0 + \lambda_1 + 2) \ln 2 }.
\end{eqnarray}

\vspace*{-22mm}\par
\begin{center}
\mbox{ }\hspace*{-10mm}
\epsfysize=160mm
\epsfbox{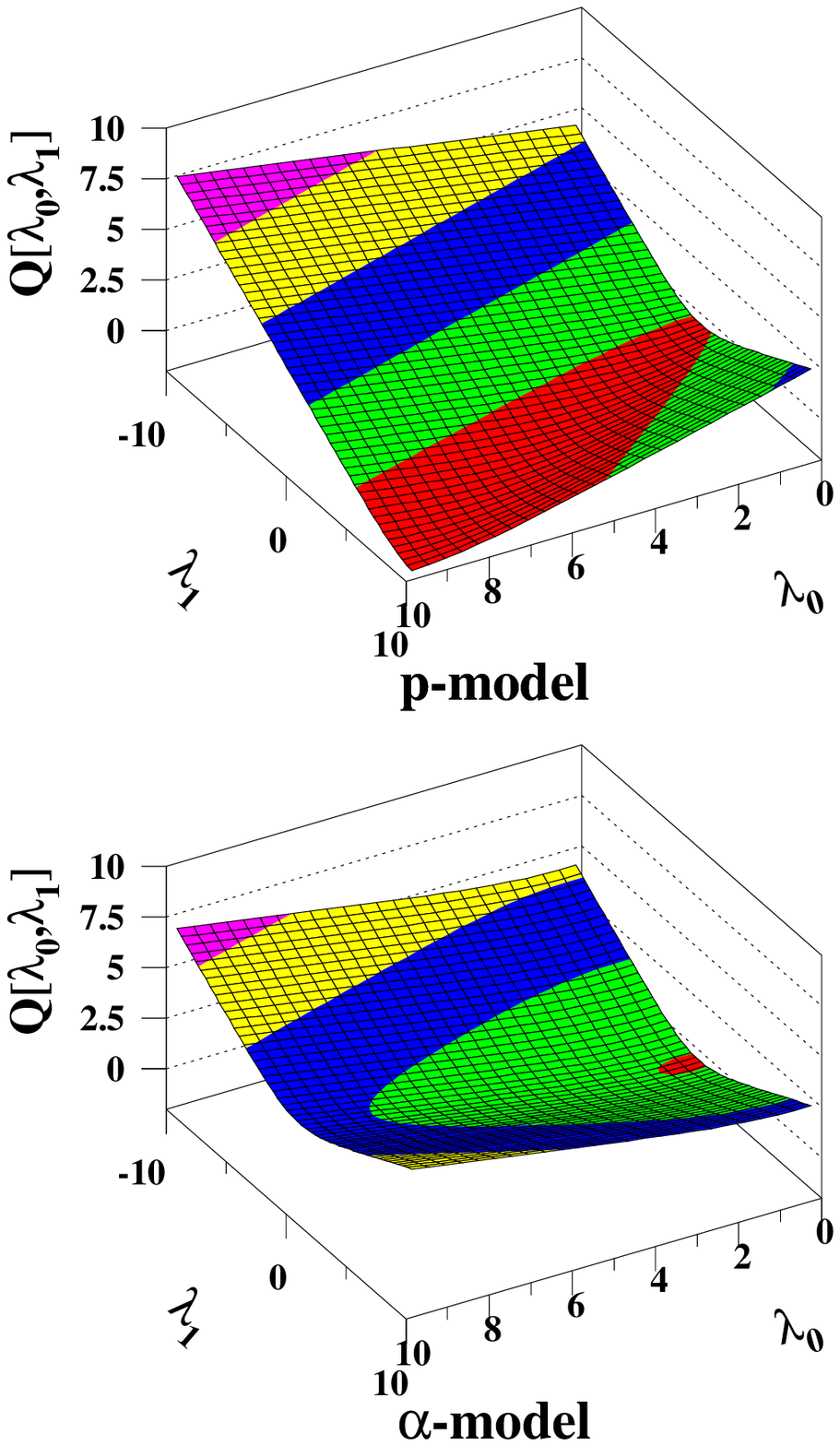}
\end{center}
\vspace*{-25mm}\par
{\noindent\bf Figure 2:}\\
Branching generating function $Q[\lambda_0, \lambda_1]$
for the $p$- and $\alpha$-models for $\alpha = 0.4$,
shown for $0 \leq \lambda_0 \leq 10$; $-10 \leq  \lambda_1 \leq 10$.
The intersection of $Q$ with the $\lambda_0{=}0$ plane is related to the
multifractal $\tau(\lambda_0)$ curve through Eq.\ (\ref{zwoelf}).
Note that these curves are identical for the $\alpha-$ and $p$-models,
while otherwise the b.g.f.'s differ significantly.
\\

Of course, when confronted with real turbulence data, the proposed
inversion will not be this straightforward as the problems of
homogeneity \cite{GRE97} and statistical dependence of multipliers
\cite{NEL96,PED96,SRE95} have to be taken into account.

Within the above limitations, 
we envisage many and diverse applications of our analytic solution 
in many branches of physics. Besides fully-developed
turbulence, the case of high-energetic multiparticle
branching processes immediately comes
to mind. For the latter, 
the $\alpha$- and $p$-models have already been used in this
context as simulation toy models \cite{BIA86,LIP89}.
Implications in this and, for example, random multiplicative 
process calculations in large-scale structure formation in the universe
\cite{PAN97} remain to be explored.
\vspace*{-5mm}\par

\acknowledgements 
This work was supported in part by the South African Foundation for
Research Development. PL acknowledges support by APART of the Austrian
Academy of Sciences.

\vspace*{-5mm}\par


\begin{thebibliography}{99}


\bibitem[*]{PCnote}Dedicated to the memory of Peter Carruthers,
          teacher and friend.

\bibitem{MEN91}C.\ Meneveau and K.R.\ Sreenivasan,
         J.\ Fluid Mech.\ {\bf 224}, 429 (1991). 

\bibitem{FRI95}U.\ Frisch,
         {\it Turbulence}, (Cambridge University Press, Cambridge, 1995).

\bibitem{HAL86}T.\ Halsey, M.\ Jensen, L.\ Kadanoff, I.\ Procaccia and
         B.\ Shraiman,
         Phys.\ Rev.\ A{\bf 33}, 1141 (1986).

\bibitem{ARG90}F.\ Argoul, A.\ Arneodo, J.\ Elezgaray, G.\ Grasseau
         and R.\ Murenzi,
         Phys.\ Rev.\ A{\bf 41}, 5537 (1990).

\bibitem{WOL96}E.A.\ De Wolf, I.M.\ Dremin and W.\ Kittel,
         Phys.\ Rep.\ {\bf 270}, 1 (1996).

\bibitem{GAR96}P.\ Garrido, S.\ Lovejoy and D.\ Schertzer,
         Physica A{\bf 225}, 294 (1996).

\bibitem{FED88}J.\ Feder, 
         {\it Fractals}, (Plenum Press, New York, 1988).
             
\bibitem{MEN87}C.\ Meneveau and K.R.\ Sreenivasan,
         Phys.\ Rev.\ Lett.\ {\bf 59}, 1424 (1987).

\bibitem{SCH85}D.\ Schertzer and S.\ Lovejoy, 
        in {\it Turbulent Shear Flows 4}, 1985, edited by
        L.J.S.\ Bradbury, F.\ Durst, B.\ Launder, F.W.\ Schmidt
        and J.H.\ Whitelaw, (Springer, 1985), pp.\ 7--33.

\bibitem{CAT87}M.E.\ Cates and J.M.\ Deutsch,
        Phys.\ Rev.\ A{\bf 35}, 4907 (1987).

\bibitem{MEN90}C.\ Meneveau and A.B.\ Chhabra,
        Physica A{\bf 164}, 564 (1990).

\bibitem{GRE95}M.\ Greiner, P.\ Lipa and P.\ Carruthers,
        Phys.\ Rev.\ E{\bf 51}, 1948 (1995).

\bibitem{GRE96}M.\ Greiner, J.\ Giesemann, P.\ Lipa and P.\ Carruthers,
        Z.\ Phys.\ C{\bf 69}, 305 (1996).

\bibitem{NEL96}M.\ Nelkin and G.\ Stolovitzky,
        Phys.\ Rev.\ E{\bf 54}, 5100 (1996).

\bibitem{PED96}G.\ Pedrizzetti, E.A.\ Novikov and A.A.\ Praskovsky,
        Phys.\ Rev.\ E{\bf 53}, 475 (1996).

\bibitem{SRE95}K.R.\ Sreenivasan and G.\ Stolovitzky,
        J.\ Stat.\ Phys.\ {\bf 78}, 311 (1995).

\bibitem{NOV71}E.A.\ Novikov,
        Prikl.\ Mat.\ Mekh.\ {\bf 35}, 266 (1971);
        Phys.\ Fluids A{\bf 2}, 814 (1990).

\bibitem{GRE97}M.\ Greiner, J.\ Giesemann and P.\ Lipa,
        Phys.\ Rev.\ E{\bf 56}, 4263 (1997).

\bibitem{BIA86}A.\ Bia\l as and R.\ Peschanski,
        Nucl.\ Phys.\ B{\bf 273}, 703 (1986).

\bibitem{LIP89}P.\ Lipa and B.\ Buschbeck,
        Phys.\ Lett.\ B{\bf 223}, 465 (1989).

\bibitem{PAN97}J.\ Pando, P.\ Lipa, M.\ Greiner and L.-Z.\ Fang,
        preprint AZPH-TH/97-03, Ap.\ J.\ (in press).

\end{thebibliography}
\end{document}